\documentclass[a4paper]{jpconf}
\usepackage{graphicx}
\begin{document}
\title{$Z$-dark search with the ATLAS detector}

\author{Elizabeth Castaneda-Miranda}

\address{Department of Physics, University of Johannesburg, South Africa}

\ead{elizabeth.castaneda.miranda@cern.ch}


\begin{abstract}
The search of the ``hidden sector'' via new neutral light bosons Z-dark ($Z_{d}$) could be revealed by the study of the decay of the 
discovered Higgs-like boson or any other undiscovered Higgs boson. After the LHC concluded a successful first period of running, 
the ATLAS Collaboration published its latest results on the $H\rightarrow Z_{d}Z_{d}\rightarrow 4l$ analysis using up to 20 fb$^{-1}$ 
of integrated luminosity at $\sqrt{s}=8$~TeV. In this proceeding I present a summary of the recent results on the search of the $Z_{d}$ 
in the signature $H\rightarrow Z_{d}Z_{d}\rightarrow 4l$ with the ATLAS detector at the LHC. 
\end{abstract}


\section{Introduction}
The $H\rightarrow Z_{d}Z_{d}\rightarrow 4l$ is an interesting search compatible with one of the main decay
channels used in the discovery of the SM Higgs boson, $H\rightarrow ZZ^{*}\rightarrow 4l$~\cite{atlashiggs}. This search is motivated by many 
extensions to the Standard Model (SM) which provide a candidate for the dark force mediator to explain the astrophysical observations 
of the positron excess~\cite{pamela}, or a candidate mediator of the ``hidden or dark sector''~\cite{Wells10, Curtin5, Curtin6}. 

This document presents the results on the $Z_{d}$ search, for both model independent and dependent cases. The study is based on the 
SM $H\rightarrow ZZ^{*}\rightarrow 4l$~\cite{atlashiggs, atlashiggs2}, with objective to find a significant excess in the dilepton mass region 
$15\leq m_{Z_{d}\rightarrow ll} \leq 60$~GeV, with a Higgs mass fixed $m_{H}=125$~GeV. Otherwise, 95$\%$ upper bound are set on the main parameters of interest 
as a function of the $m_{Z_{d}}$. The information presented here is reported and published by the ATLAS Collaboration~\cite{hzdzd4l}.

\section{The ATLAS detector}
It is a multipurpose detector located at point 1 at the LHC. Generally speaking, the ATLAS detector consists of four major subsystems: Inner Detector (ID), 
Calorimeter, Muon Spectrometer (MS) and Magnet System. These subsystems are integrated with the following components: Trigger and Data Acquisition 
System and Computing System~\cite{atlasperf, atlastrig}. The coordinate system used in ATLAS is the right-handed with origin at interaction point in 
the center of the detector, the z-axis along the beam line and the x-y plane perpendicular to the beam line. Cylindrical coordinates ($r,\phi$) are used in the 
transverse plane, $\phi$ is the azimuthal angle around the beam line. Observables labelled ``transverse'' are projected into the x-y plane. The pseudorapidity is
 defined in terms of the polar angle $\theta$ as $\eta=-\ln{\tan{\theta/2}}$. 

\section{Signal and background simulation samples}
The new neutral light boson is produced via the Higgs boson production ($H\rightarrow Z_{d}Z_{d}\rightarrow4l$) in the gluon fusion (ggF).
The benchmark model used is the hidden Abelian Higgs model (HAHM)~\cite{Wells10, Curtin5, Curtin6}. The samples are generated for $m_{H}=125$~GeV and 
$15\leq m_{Z_{d}}\leq60$~GeV, in steps of $5$~GeV, using {\scriptsize MadGraph} with {\scriptsize CTEQ6L1} for the parton distribution functions 
(PDF). For taking into account the parton showering, hadronization, and initial- and final-state radiation {\scriptsize PYTHIA}8 and {\scriptsize PHOTOS} are used. In the $H\rightarrow Z_{d}Z_{d}\rightarrow4l$ search, the $h\rightarrow 4l$ is normalized  from data.

The background processes considered for this study are listed in Table~\ref{MCsamples}. 
\begin{table}
\caption{\label{MCsamples} List of the background processes and their corresponding event generators, PDF's calculators, 
and their QCD and EW applied corrections. The references of the generators, used for this study, are found in~\cite{hzdzd4l}.}
\scriptsize
\begin{center}
\begin{tabular}{lllll}
\br
Process&Generator&PDF&QCD corr.&EW corr.\\
\mr
ggFF, VBF&{\scriptsize POWHEG$+$PYTHIA8}&CT10&NLO and NNLO&NLO\\
WH, ZH&{\scriptsize PYTHIA8}&CT10&NNLO&NLO\\
$t\bar{t}$H&{\scriptsize PYTHIA8}&CTEQ6L1&NLO&---\\
$q\bar{q}\rightarrow$ ZZ$^{*}$&{\scriptsize POWHEG$+$PYTHIA8}&CT10&NLO&---\\
gg $\rightarrow$ ZZ$^{*}$&{\scriptsize gg2ZZ$+$JIMMY}&CT10&NLO&---\\
Z$+$jets&{\scriptsize ALPGEN}&CTEQ6L1&NLO and NNLO&---\\
$t\bar{t}$&{\scriptsize MC@NLO$+$HERWIG, JIMMY}&CT10&NLO&---\\
WZ and WW&{\scriptsize SHERPA}&CT10&NLO&---\\
Z$J/\psi$, Z$\Upsilon$ &{\scriptsize PYTHIA8}&CTEQ6L1&---&---\\
\br
\end{tabular}
\end{center}
\end{table}
For the SM Higgs boson in ggF production, the QCD soft-gluon resummations, calculated in the next-to-next-to-leading-logarithmic 
(NNLL) approximation, are applied. The $Z+$jets production, modelled with up to five partons, is divided in two sources: $Z+$light-jets and $Zb\bar{b}$.
For data and simulation comparison, the QCD cross-section calculations (see Table~\ref{MCsamples}) are used to normalize the simulation 
for inclusive $Z$ boson and $Zb\bar{b}$ production. The $ZJ/\psi$ and $Z\Upsilon$ Monte Carlo (MC) samples are normalized using the ATLAS measurement described in~\cite{jpsi}. 
The SM Higgs productions, SM $ZZ^{*}$ and diboson processes are normalized with the theoretical cross-sections and decay branching ratios as well as their uncertainties.
For taking into account the different conditions in the pileup, as a function of the instantaneous luminosity, the simulated minimun-bias overlayed events,
generated with {\scriptsize PYTHIA8} onto the hard-scattering process, were used. Such events were reweighted according to the distribution of the 
mean number of interactions observed in data. All the MC background samples generated are reconstructed using the ``full detector simulation'' ATLAS
software~\cite{fullsim} which is based on {\scriptsize GEANT4}~\cite{geant}. The signal samples are reconstructed with the ``fast detector simulation'' 
ATLAS software~\cite{atlfast} which only parametrized the response of the electromagnetic and hadronic showers in the ATLAS calorimeter, and the rest of the 
systems are reconstructed as ``full detector simulation''.


\section{Event selection}
For this search, 20.3 fb$^{-1}$ of integrated luminosity is used which was recorded in the optimal function of the detector. All the MC processes are normalized with integrated luminosity mentioned. The event selection applied is the same for both data and MC. There are three categories that are studied: 
$4-$electrons ($4e$), $4-$muons ($4\mu$) and $2-$electrons $2-$muons ($2e2\mu$). 

It is requested that the (data/MC) events pass the combination of single-lepton and dilepton triggers. Table~\ref{trigger} shows the list of single-lepton
and dilepton triggers used, with their respective thresholds in transverse energy/momentum. All these events must contain a reconstructed primary 
vertex, defined as the vertex with highest $\sum p_{T}^{2}$ of the associated tracks, with at least three tracks of $p_{T}>0.4$~GeV each one.
\begin{table}
\caption{\label{trigger} Triggers used with their respective $E_{T}$~or~$p_{T}$ thresholds.}
\scriptsize
\begin{center}
\begin{tabular}{lll}
\br
Trigger&$E_{T}$~[GeV]&$p_{T}$~[GeV]\\
\mr
single-electron&25&---\\
single-muon&---&24\\
dielectron&12&---\\
dimuon (symm.)&---&13\\
dimuon (asymm.)&---&18 and 8\\
electron-muon&12 or 24&8\\
\br
\end{tabular}
\end{center}
\end{table}
The muon candidates selected are formed by matching reconstructed ID tracks with either a complete track or track-segment reconstructed in MS
\cite{muon}. The acceptance is extended using tracks reconstructed in the forward region of the MS $(2.5<|\eta|<2.7)$, which is outside 
the ID coverage. If both an ID or a complete MS track are present, the two independent momentum measurements are combined; otherwise the 
information of the ID or the MS is used alone. In the case of the electrons candidates, they must have well-reconstructed ID track pointing to an 
electromagnetic calorimeter cluster and the cluster should satisfy a set of identification criteria~\cite{electron}. Tracks associated with 
electromagnetic clusters are fitted using a Gaussian-Sum Filter~\cite{electron}, which allows for bremsstrahlung energy losses to be taken 
into account.
Each electron (muon) must satisfy $p_{T}>7$~GeV ($p_{T}>6$~GeV) and be measured in the range $|\eta|<2.47$ ($|\eta|<2.7$). The
longitudinal impact parameters of the leptons along the beam axis $z_{0}$ are required to be within 10 mm of the reconstructed primary vertex (muons 
without ID track are exempted of this requirement). To reject the cosmic rays, muon tracks are required to have a transverse impact parameter $d_{0}$ 
(closest approach to the primary vertex in the x-y plane) of less than 1mm.
For avoiding double-counting of leptons, an overlap removal is applied. If two electrons share the same ID track or within in a cone 
($\Delta R\equiv \sqrt{(\Delta\phi)^{2}+(\Delta\eta)^{2}}<0.1$), the one with the highest transverse energy deposit in the calorimeter ($E_{T}$) is kept. 
An electron within in a muon cone ($\Delta R=0.2$) is removed, and a calorimeter-based reconstructed muon within an electron cone ($\Delta R=0.2$) 
is removed.

\begin{figure}[h]
\begin{center}
\begin{minipage}{16pc}
\includegraphics[width=18.5pc]{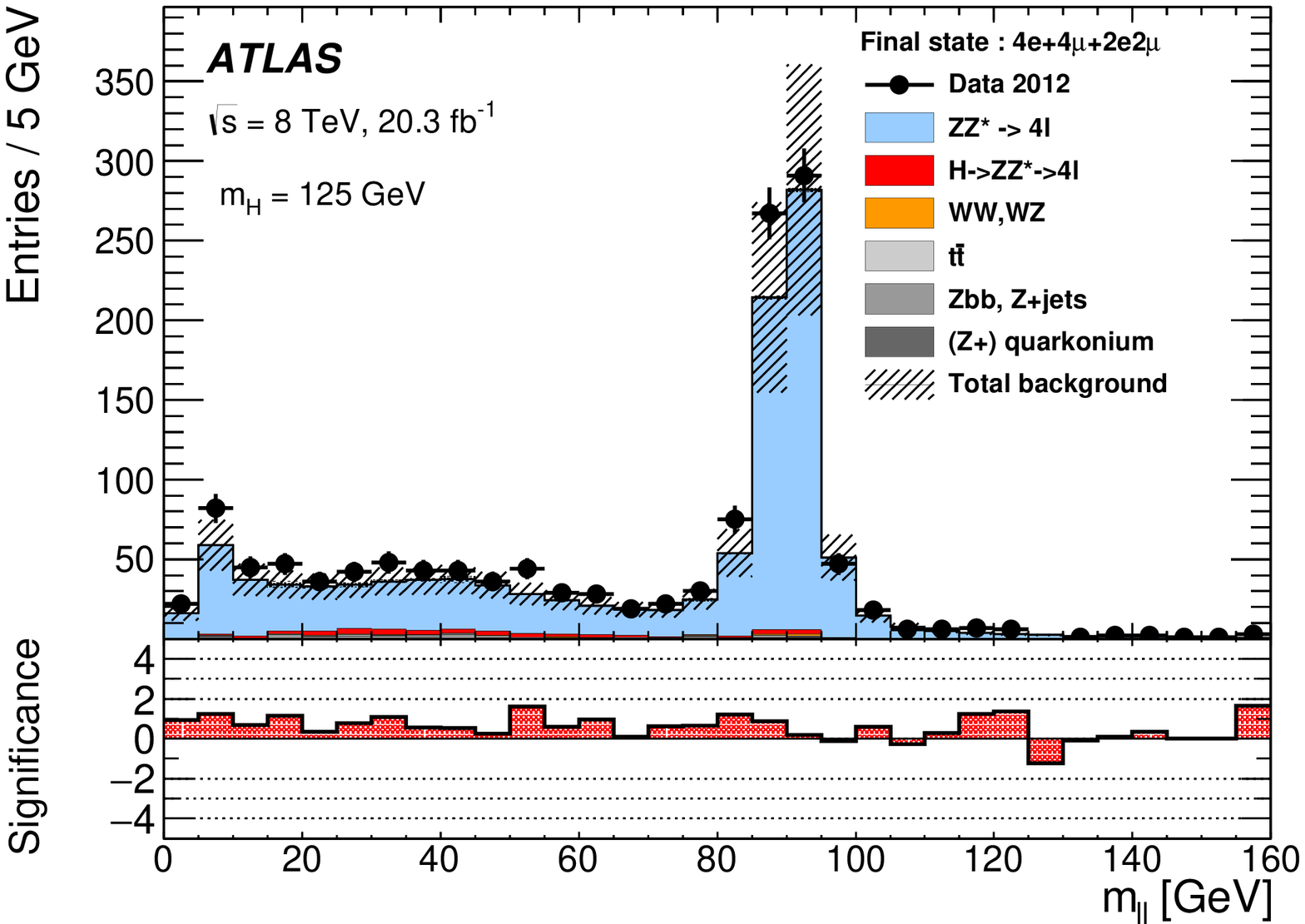}
\caption{\label{dimassIP} Dilepton invariant mass $m_{ll=12, 34}$ for $m_{H}=125$~GeV at the IP requirements.}
\end{minipage}\hspace{2pc}%
\begin{minipage}{17.5pc}
\includegraphics[width=19pc]{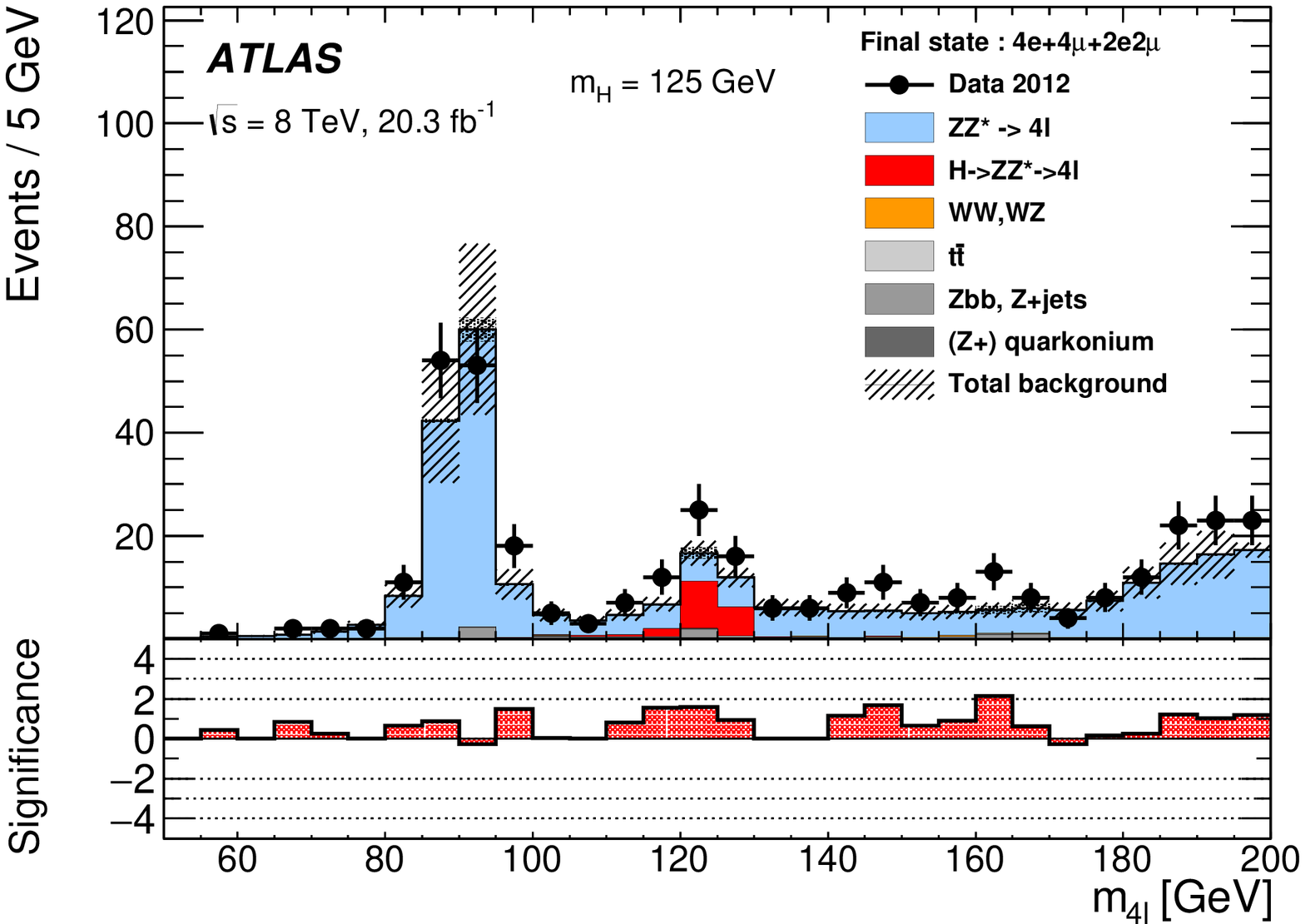}
\caption{\label{fourmassIP} Four-lepton invariant mass for $m_{H}=125$~GeV at the IP requirements.}
\end{minipage}
\end{center} 
\end{figure}

After selecting the leptons with proper quality, events with at least four leptons, with all possible combinations of four leptons containing two same-flavor and 
opposite-sign (SFOS) defined the ``quadruplets'', are kept. Each lepton is ordered by $p_{T}$, the three highest-$p_{T}$ leptons should have: 
$p_{Tl1}>20$~GeV, $p_{Tl2}>15$~GeV and $p_{Tl3}>10$~GeV. It is also required that at least one (two) of the lepton(s), from the quadruplet, 
must satisfy the single-lepton (dilepton) trigger requirements. Each dilepton invariant mass or each SFOS pair, from the quadruplet, are denoted by 
$m_{12}$ (formed by leptons 1 and 2) and $m_{34}$ (formed by leptons 3 and 4). For the $H\rightarrow Z_{d}Z_{d}\rightarrow 4l$ study, both $Z_{d}$ boson 
are considered to be on-shell, because of that, there is not any distinction between the SFOS pairs. If in the event there are several ``quadruplets'', only the
quadruplet that satisfies the minimum value of $\Delta m \equiv |m_{12} - m_{34}|$ is selected. Subsequently, the transverse impact parameter significance (IP) and the isolation requirements are imposed to the leptons of the quadruplet selected, and four final requirements that defined the two
signal regions (SR) considered in this study are shown in Table~\ref{signalregions}. In Figs.~\ref{dimassIP} and ~\ref{fourmassIP} show the dilepton and 
four-lepton invariant masses after the IP requirements, respectively, in the dilepton mass the $m_{12}$ and $m_{34}$ are combined.
\begin{table}
\caption{\label{signalregions} Requirements after selecting the one quadruplet, and the loose and tight SR definitions.}
\scriptsize
\begin{center}
\begin{tabular}{ll}
\br
Requirements\\
\mr
Track isolation& $\frac{\sum p_{T}^{tracks} }{p_{T}^{lepton}} (\Delta R = 0.2) < 0.15$ \\
Calorimeter isolation&$\frac{\sum E_{T}^{clusters}}{ E_{T}^{e}(p_{T}^{\mu})} (\Delta R = 0.2) <$0.2 (0.3 and $\mu$'s no ID 0.15)\\
IP& $\frac{d_{0}}{\sigma_{0}^{IP~uncer.}} < 6.5$ for $e$'s (3.5 for $ \mu$'s) \\
Loose SR &$115<m_{4}<130$~GeV, $|m_{Z_{d}}-m_{ll}|<\frac{m_{H}}{2},~m_{H}=125$ GeV, \\ 
& and $Z$, $J/\psi-\Upsilon$ vetoes: $|m_{Z_{PDG}}-m_{ll}|<10$ GeV, \\
&$m_{ll}<12$~GeV (ll=12,34, and 23, 14)\\
Tight SR& Loose SR and $|m_{Z_{d}}-m_{ll}|<\delta m$ \\ 
&(5/3/4.5 GeV for $4e/4\mu/2e2\mu$)\\
\br
\end{tabular}
\end{center}
\end{table}

\section{Background estimation}\label{bckgrd}
\begin{figure}[h]
\begin{center}
\begin{minipage}{16pc}
\includegraphics[width=18.5pc]{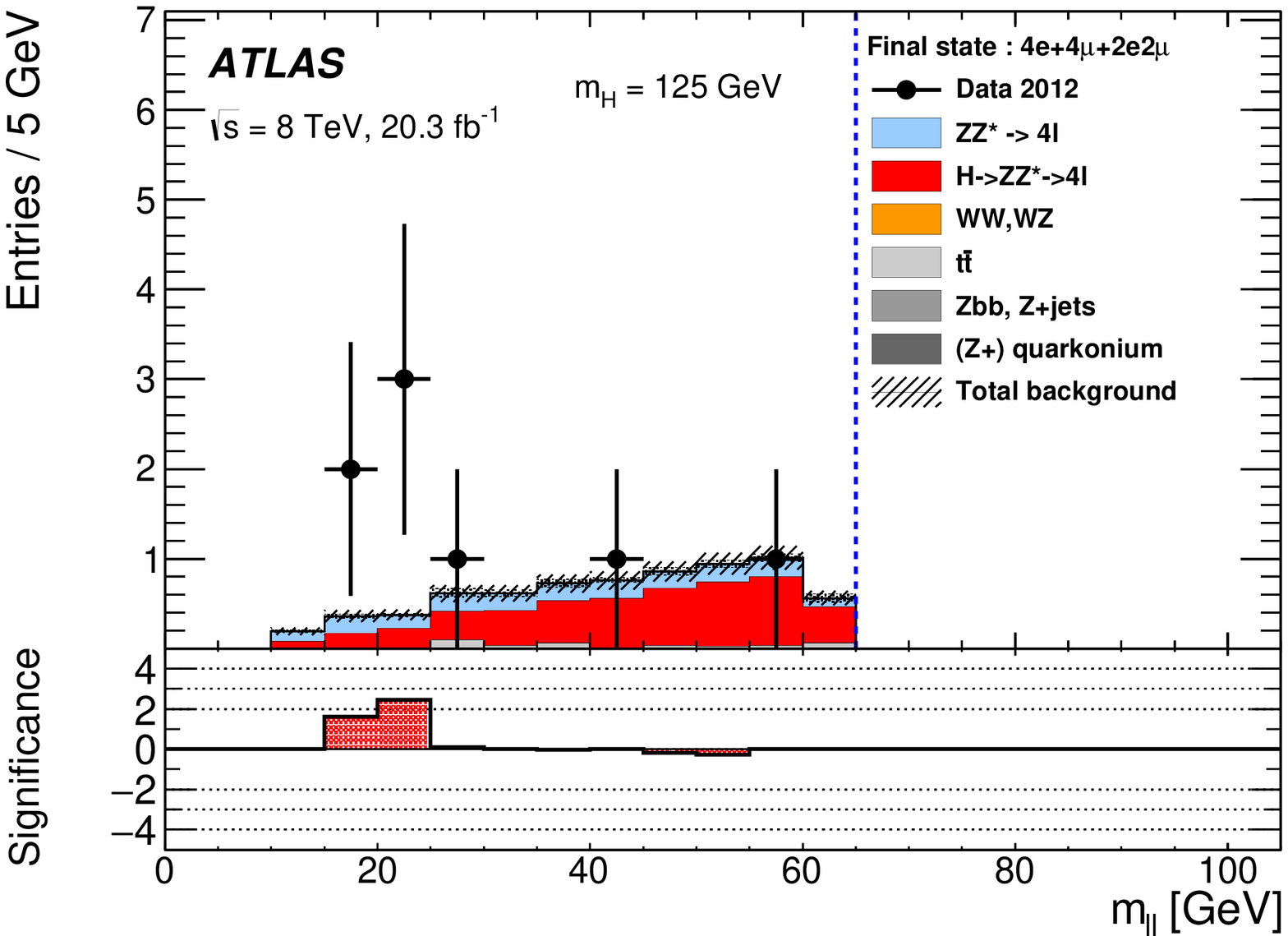}
\caption{\label{dimassTSR}Dilepton invariant mass $m_{ll=12, 34}$ for $m_{H}=125$~GeV at the Loose SR. }
\end{minipage}\hspace{2pc}%
\begin{minipage}{16.5pc}
\includegraphics[width=18pc]{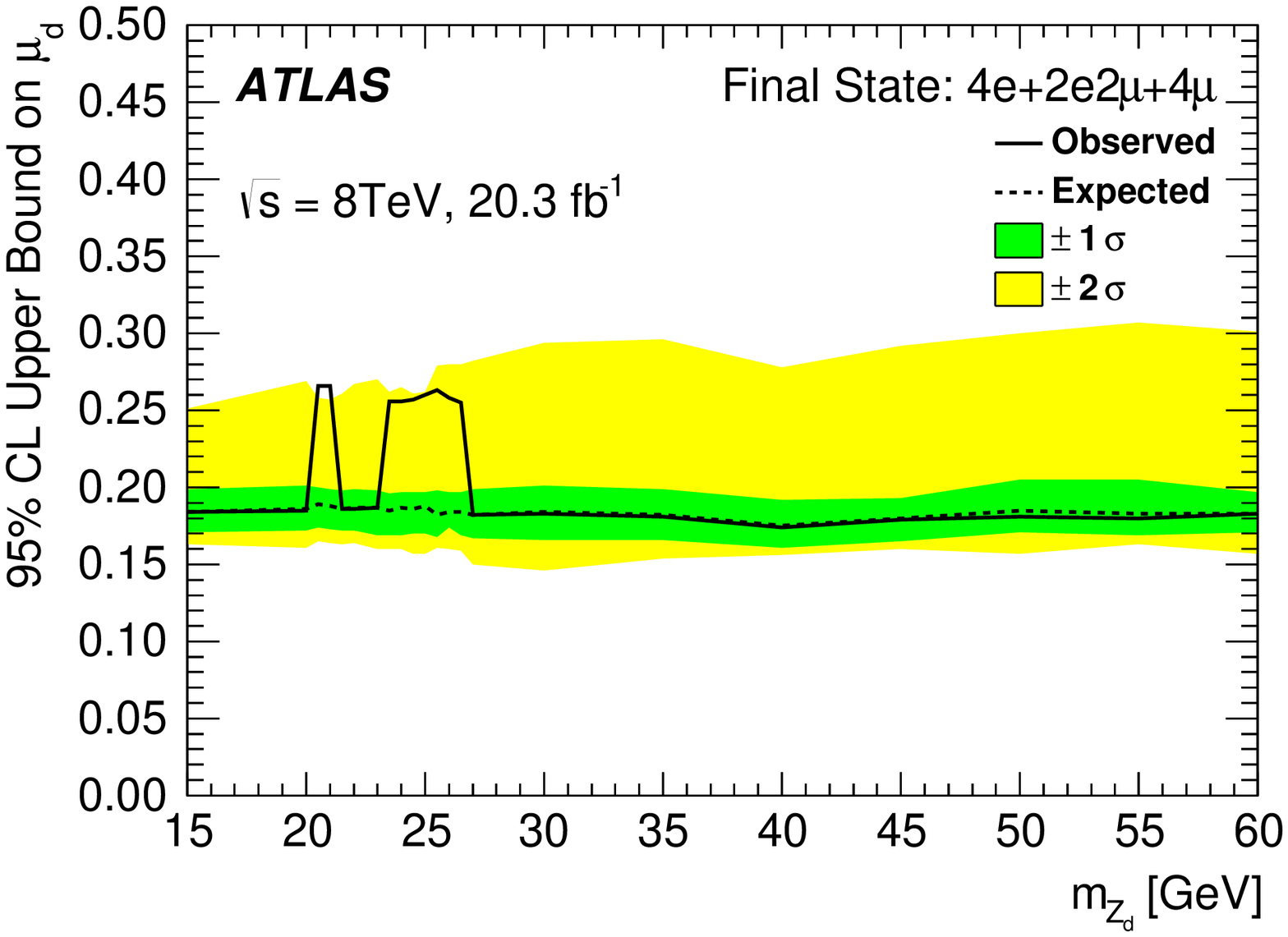}
\caption{\label{mulimits} The 95$\%$ C.L. upper bound on the signal strength $\mu_{d}$, see Eq~\ref{Eq1}.}
\end{minipage}
\end{center} 
\end{figure}

For the  $H\rightarrow Z_{d}Z_{d}\rightarrow 4l$ analysis, the main background contributions are $ZZ^{*}\rightarrow 4l$ and 
$H\rightarrow ZZ^{*}\rightarrow 4l$ which are reduced considerably after applying the tight SR (see Table~\ref{signalregions}). Also small contribution from 
$Z+$jets, $t\bar{t}$, $WW$ and $WZ$ can be seen from Fig.~\ref{dimassTSR}. The $H\rightarrow ZZ^{*}\rightarrow 4l$, $ZZ^{*}\rightarrow 4l$, $WW$ and 
$WZ$ background processes are estimated from MC simulation, their normalization is based on theoretical cross-sections calculations, integrated luminosity 
recorded, and acceptance efficiency. In the case of zero MC background expected events, in the tight SR, 68$\%$ C.L. upper bound, which corresponds to 1.14 
events~\cite{pdg}, is estimated as $N_{background} < L \times \sigma \times \frac{1.14}{N_{total}}$,
where $L$ is the integrated luminosity recorded, $\sigma$ the cross-section of the background processes and $N_{tot}$ is the total number of
weighted events simulated for the background processes. This background estimation is validated using a control region defined by reversing, the four-lepton 
invariant mass requirement ($m_{4l}<115$~GeV or $m_{4l}>115$~GeV), good agreement is found between the expectation and observation results~\cite{hzdzd4l}.

\section{Systematic uncertainties}\label{Secsys}
The systematic uncertainties on the theoretical cross-section calculations as well as the detector uncertainties are taken into account. The PDF effects,
$\alpha_{s}$, renormalization and factorization scale uncertainties are applied to the $ZZ^{*}$ backgrounds, and for all the SM Higgs productions applied on the total
inclusive cross-sections, using values obtained from~\cite{atlashiggs3}. Uncertainties due to the limited number of MC events in the $t\bar{t}$, $Z+$jets, $ZJ/\psi$,
$Z\Upsilon$ and $WW/WZ$ background processes are estimated (see Sec.~\ref{bckgrd}). The luminosity uncertainty~\cite{lumi} is applied to all signal and
background yields. The detector systematic uncertainties on the electron and muons, within the acceptance on SR, are applied (see Table~\ref{sys}): 
uncertainties on electron/muon identification, reconstruction and trigger efficiency (summarized in Electron/Muon identification), on the electron/muon 
energy/momentum scale, and on the electron/muon energy/momentum resolution (summarized in Electron/Muon energy/momentum scale). Also the 
uncertainties from the isolation and IP requirements are taken into account. 
\begin{table}
\caption{\label{sys} Relative systematic uncertainties on the event yields.}
\scriptsize
\begin{center}
\begin{tabular}{llll}
\br
Source&$4\mu$&$4e$&$2e2\mu$\\
\mr
Electron identification&---&6.7&3.2\\
Electron energy scale&---&0.8&0.3\\
Muon identification&2.6&---&1.3\\
Muon momentum scale&0.1&---&0.1\\
Luminosity&2.8&2.8&2.8\\
ggF QCD&7.8&7.8&7.8\\
ggF PDF's and $\alpha_{s}$&7.5&7.5&7.5\\
ZZ$^{*}$ Normalization&5.0&5.0&5.0\\
\br
\end{tabular}
\end{center}
\end{table}


\section{Results}
\begin{figure}[h]
\begin{center}
\begin{minipage}{15pc}
\includegraphics[width=17pc]{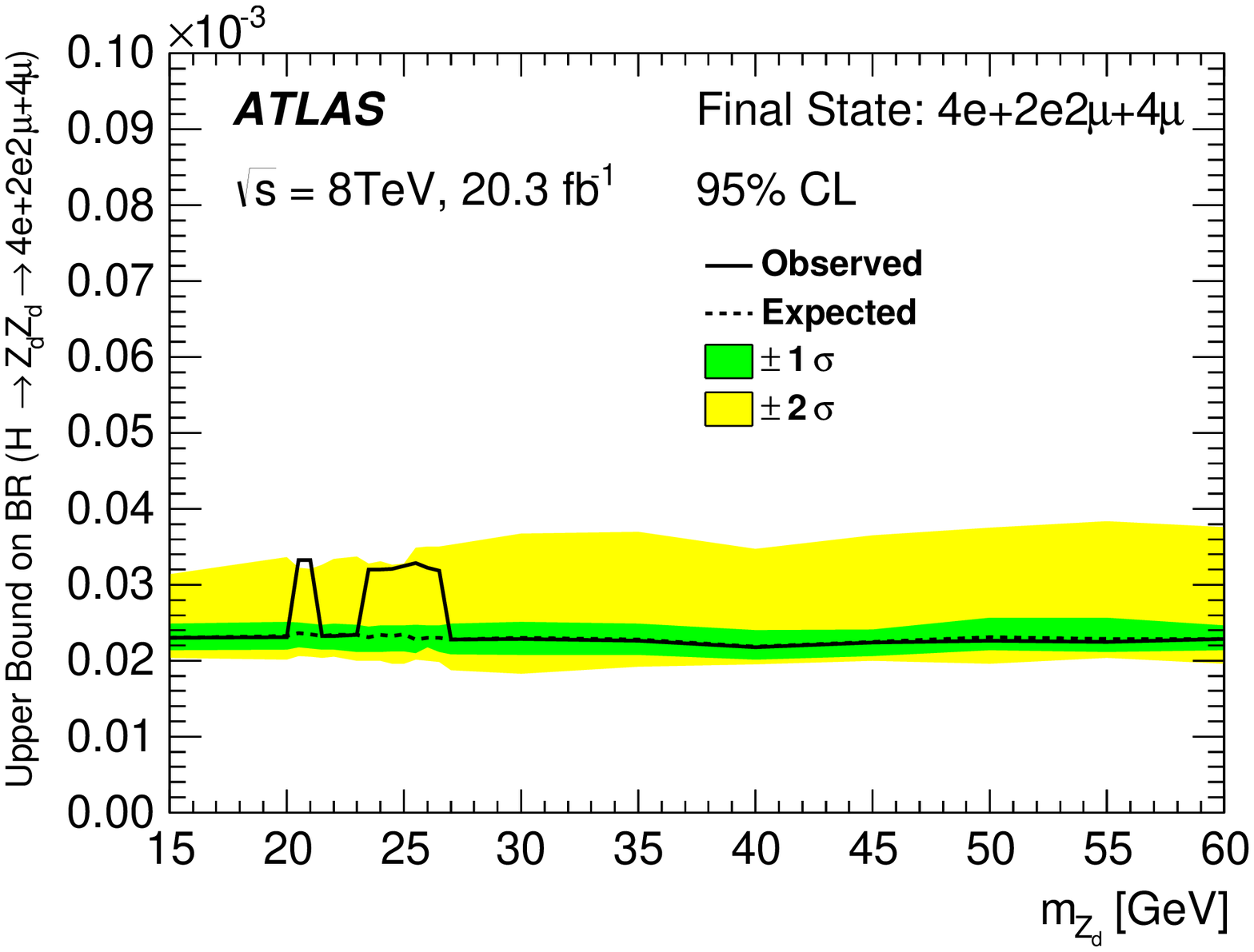}
\caption{\label{BRpic} The 95$\%$ C.L. upper bound on the $BR(H\rightarrow Z_{d}Z_{d}\rightarrow 4l)$.}
\end{minipage}\hspace{3.1pc}%
\begin{minipage}{16.5pc}
\includegraphics[width=17pc]{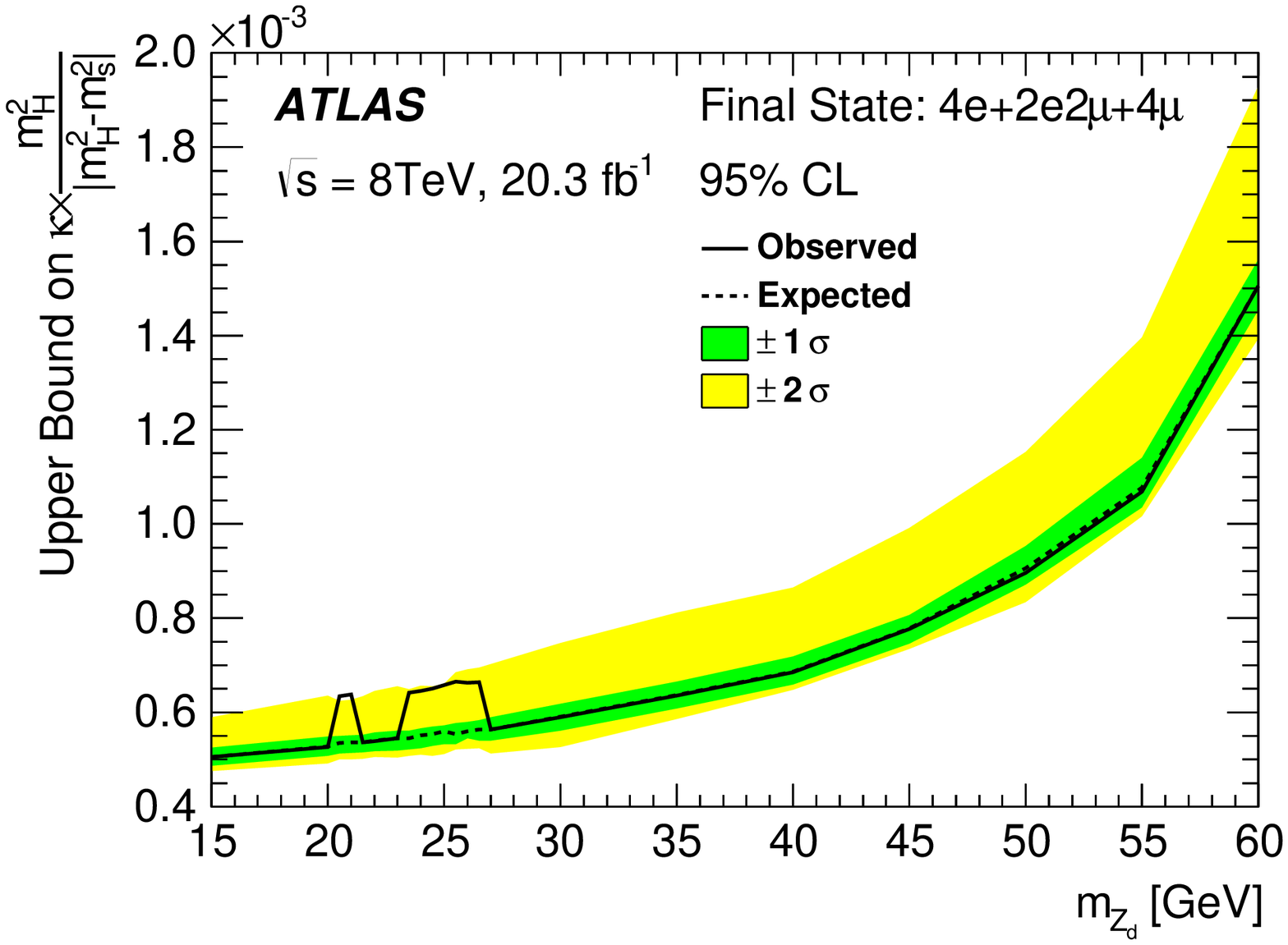}
\caption{\label{ModDep} The 95$\%$ C.L. upper bound on the Higgs mixing parameter, see Eq.~\ref{kapa}. }
\end{minipage}
\end{center} 
\end{figure}

Four data events pass after applying the loose SR requirements: one in the $4e$, two in the $4\mu$ and one in the $2e2\mu$ channel. Only two of these 
events pass the tight SR: the event in the $4e$ and one of the events in the $4\mu$ channel. The events in $4e$ and $4\mu$ channels are consistent with 
the $Z_{d}$ mass range $23.5\leq m_{Z_{d}} \leq 26.5$~GeV and $20.5\leq m_{Z_{d}} \leq 21.0$~GeV, respectively, as shown in Fig.~\ref{dimassTSR}. 
In the mass range of 15 to 60 GeV the interpolated histograms are used~\cite{interpol}, generated, based on the MC signal simulations, 
in steps of 0.5 GeV to obtain the signal acceptance at the hypothesized $m_{Z_{d}}$. Table~\ref{results} shows the expected and observed number
of events after applying the tight SR requirements.
\begin{table}
\caption{\label{results} The expected and observed number of events in the tight signal region for each final state are shown. The results are 
for the mass regions $m_{Z_{d}}=25$~and~20.5~GeV. The statistical and systematic uncertainties are given for the signal and 
background expected events. The $H\rightarrow ZZ^{*}\rightarrow 4l$ expected events are the combination of the ggF, VBF, $ZH$, $WH$ 
and $t\bar{t}H$ processes.}
\scriptsize
\begin{center}
\lineup
\begin{tabular}{llll}
\br
Process&$\0\0\0\0\0\0\0\04e$&\0\0\0\0\0\0\0\0$4\mu$&\0\0\0\0\0\0\0$2e2\mu$\\
\mr
$H\rightarrow ZZ^{*}\rightarrow 4l$       & $(1.5\pm 0.3 \pm 0.2)\times 10^{-2}$ & $(1.0\pm 0.3 \pm 0.3)\times 10^{-2}$&$(2.9\pm 1.0 \pm 2.0)\times 10^{-3}$\\
$ZZ^{*}\rightarrow 4l$                            & $(7.1\pm 3.6 \pm 0.5)\times 10^{-4}$ & $(8.4\pm 3.8 \pm 0.5)\times 10^{-3}$ & $(9.1\pm 3.6 \pm 0.6)\times 10^{-3}$\\
$WW$, $WZ$                                        &\0\0\0\0$<0.7\times 10^{-2}$ 	&\0\0\0\0 $<0.7\times 10^{-2}$	&\0\0\0\0$<0.7\times 10^{-2}$\\
$t\bar{t}$                                               &\0\0\0\0$<3.0\times 10^{-2}$          	&\0\0\0\0 $<3.0\times 10^{-2}$		&\0\0\0\0$<3.0\times 10^{-2}$\\
$Zbb$, $Z+$jets                                   &\0\0\0\0$<0.2\times 10^{-2}$          	&\0\0\0\0 $<0.2\times 10^{-2}$		&\0\0\0\0$<0.2\times 10^{-2}$\\
$ZJ/\psi$, $Z\Upsilon$                         &\0\0\0\0$<2.3\times 10^{-3}$   		&\0\0\0\0 $<2.3\times 10^{-3}$		&\0\0\0\0$<2.3\times 10^{-3}$\\
Total background                                &\0\0\0\0$<5.6\times 10^{-2}$ 			&\0\0\0\0 $<5.9\times 10^{-2}$		&\0\0\0\0$<5.3\times 10^{-2}$\\
Data                                                   &\0\0\0\0\0\0\0\0\01					&\0\0\0\0\0\0\0\0\00				&\0\0\0\0\0\0\0\0\00\\
\mr
$H\rightarrow ZZ^{*}\rightarrow 4l$    & $(1.2\pm 0.3 \pm 0.2)\times 10^{-2}$ 		& $(5.8\pm 2.0 \pm 2.0)\times 10^{-3}$	&$(2.6\pm 1.0 \pm 0.2)\times 10^{-3}$\\
$ZZ^{*}\rightarrow 4l$                        & $(3.5\pm 2.0 \pm 0.2)\times 10^{-3}$ 		& $(4.1\pm 2.7 \pm 0.2)\times 10^{-3}$ 	& $(2.0\pm 0.6 \pm 0.1)\times 10^{-2}$\\
$WW$, $WZ$                                   &\0\0\0\0$<0.7\times 10^{-2}$ 					&\0\0\0\0 $<0.7\times 10^{-2}$				&\0\0\0\0$<0.7\times 10^{-2}$\\
$t\bar{t}$                                           &\0\0\0\0$<3.0\times 10^{-2}$ 					&\0\0\0\0 $<3.0\times 10^{-2}$				&\0\0\0\0$<3.0\times 10^{-2}$\\
$Zbb$, $Z+$jets                               &\0\0\0\0$<0.2\times 10^{-2}$ 					&\0\0\0\0 $<0.2\times 10^{-2}$				&\0\0\0\0$<0.2\times 10^{-2}$\\
$ZJ/\psi$, $Z\Upsilon$                     &\0\0\0\0$<2.3\times 10^{-3}$ 					&\0\0\0\0 $<2.3\times 10^{-3}$				&\0\0\0\0$<2.3\times 10^{-3}$\\
Total background                             & \0\0\0\0$<5.3\times 10^{-2}$ 					& \0\0\0\0$<5.1\times 10^{-2}$				&\0\0\0\0$<6.4\times 10^{-2}$\\
Data                                                 &\0\0\0\0\0\0\0\0\00							&\0\0\0\0\0\0\0\0\01						&\0\0\0\0\0\0\0\0\00\\
\br
\end{tabular}
\end{center}
\end{table}
In the absence of any significant excess, the 95$\%$ C.L. upper bound is computed on the parameter of interest, the signal strength, which is defined as: 
\begin{equation}\label{Eq1}
\mu_{d}=\frac{\sigma\times BR(H\rightarrow Z_{d}Z_{d}\rightarrow 4l)}{[\sigma\times BR(H\rightarrow ZZ^{*}\rightarrow 4l)]_{SM}} .
\end{equation}
The limits are computed for each final state and their combination for the model independent case, following the $CL_{s}$ modified frequentist 
formalism with the profile-likelihood test statistic~\cite{CLS}. The nuisance parameters associated with the systematic uncertainties described 
in Sec~\ref{Secsys} are profiled. The systematic uncertainties on Table~\ref{sys} are 100$\%$ correlated between the signal and background. 
In Figs.~\ref{mulimits} and~\ref{BRpic} show the 95$\%$ C.L. upper bound on the $\mu_{d}$ and on the branching ratio of $H\rightarrow Z_{d}Z_{d} \rightarrow 4l$ 
as a function of the mass $m_{Z_{d}}$ for the combination of the three final states: $4e$, $4\mu$ and $2e2\mu$; assuming the SM Higgs boson cross-section and
$BR(H\rightarrow ZZ^{*}\rightarrow 4l)_{SM}=1.25\time 10^{-4}$~\cite{atlashiggs3}.

For the model dependent case, the simplest benchmark model is the SM plus a dark vector boson and a dark Higgs boson~\cite{Wells10, Curtin5, Curtin6}, where the 
branching ratio of the $Z_{d} \rightarrow ll$ is given as a function of the $m_{Z_{d}}$, which can be used to convert the upper bound measurement of the 
$\mu_{d}$ into an upper bound on the branching ratio BR($H\rightarrow Z_{d}Z_{d}$), assuming the SM Higgs production cross-section~\cite{hzdzd4l}. The
$H\rightarrow Z_{d}Z_{d}$ decay can be used to obtain an $m_{Z_{d}}-$dependent limit on an Higgs mixing parameter $\kappa'$~\cite{Curtin5}:
\begin{equation}\label{kapa}
\kappa'=\kappa\times\frac{m^{2}_{H}}{|m_{H}^{2}-m_{s}^{2}|} ,
\end{equation}
where $\kappa$ is the size of the Higgs portal coupling and $m_{s}$ is the mass of the dark Higgs boson. The partial width of the $H\rightarrow Z_{d}Z_{d}$ 
is given in terms of $\kappa$~\cite{Curtin5}. In the regime where the Higgs mixing parameter dominates ($\kappa \gg \epsilon$), $m_{s}>m_{H}/2$, 
$m_{Z_{d}}<m_{H}/2$ and $H\rightarrow Z_{d}Z^{*}\rightarrow 4l$ are negligible, the only relevant decay is $H\rightarrow Z_{d}Z_{d}$. The Higgs portal 
coupling parameter $\kappa$ is~\cite{Curtin5}:
\begin{equation}
\kappa^{2}=\frac{\Gamma_{SM}}{f(m_{Z_{d}})}\frac{BR(H\rightarrow Z_{d}Z_{d})}{1-BR(H\rightarrow Z_{d}Z_{d})}.
\end{equation}
The upper bound on the effective Higgs mixing parameter as a function of $m_{Z_{d}}$ is shown in Fig.~\ref{ModDep} for $m_{H}/2<m_{s}<2m_{H}$, 
this corresponds to an upper bound on the Higgs portal coupling in the range $\kappa \sim (1-10)\times 10^{-4}$.
\section{Conclusions}
The search of the $Z_{d}$ light boson in a mass range of $15\leq m_{Z_{d}} \leq60$~GeV, via $H\rightarrow Z_{d}Z_{d}\rightarrow 4l$ with $m_{H}=125$~GeV, 
is presented at $\sqrt{s}=8$~TeV. Two data events in the tight SR are observed in $4e$ and $4\mu$ final states: $4e$, $m_{Z_{d1,2}} = 21.8,~28.1$ GeV and $4\mu$,  
$m_{Z_{d1,2}} = 23.2,~18.0$ GeV. Since there is not significant excess observed, the 95$\%$ C.L. upper bound on the $\mu_{d}$, and on the $BR(H\rightarrow Z_{d}Z_{d}\rightarrow 4l)$
are set, combining the three final states. For the mass range $20\leq m_{Z_{d}}\leq28$~GeV, in the model independent case, the upper 
bounds for $\mu_{d}$ are $(18-27)\times 10^{-2}$; in case of the model depending study, the upper bound for the  $BR(H\rightarrow Z_{d}Z_{d}\rightarrow 4l)$ is 
$(2.1-3.2)\times 10^{-5}$, and the upper bound on the Higgs mixing parameter is $(5.0-6.5)\times 10^{-4}$ (see Fig.~\ref{mulimits}, \ref{BRpic} and~\ref{ModDep}).

\subsection{Acknowledgments}
It is a pleasure to thank the ATLAS Collaboration, to CONACyT for the support received as a postdoctoral fellow with the University of Johannesburg, 
to Professor Simon Connell, Ketevi Assamagan and the University of Cape Town for giving me the opportunity to be part of the SouthAfrican group in the ATLAS Collaboration 
and for working in this project.

\end{document}